\documentstyle{article}

\def\abstract#1{\vskip 7mm 
        \begin{center}{\large Abstract}\par \smallskip
                \begin{minipage}[c]{12cm}
                        \small #1
                \end{minipage}
        \end{center}
}
\def\title#1{\begin{center}{\Large\bf #1}\end{center}}
\def\author#1{\vskip 5mm \begin{center}{#1}\end{center}}
\def\address#1{\begin{center}{\it #1}\end{center}}
\makeatletter
\@ifundefined{lesssim}{\def\lesssim{\mathrel{\mathpalette\vereq<}}}{}
\@ifundefined{gtrsim}{\def\gtrsim{\mathrel{\mathpalette\vereq>}}}{}
\def\vereq#1#2{\lower3pt\vbox{\baselineskip1.5pt \lineskip1.5pt
\ialign{$\m@th#1\hfill##\hfil$\crcr#2\crcr\sim\crcr}}}
\makeatother

\begin{document}

\title{%
  Thermodynamics of Near-Extreme Black Holes
}
\author{%
  Don N. Page\footnote{E-mail:don@phys.ualberta.ca}
}
\address{%
  CIAR Cosmology Program \\
  Department of Physics \\
  University of Alberta \\
  Edmonton, Alberta \\
  Canada T6G 2J1
}

\abstract{
	The thermodynamics of nearly-extreme charged
black holes depends upon the number of ground states
at fixed large charge and upon the distribution of excited
energy states.  Here three possibilities are examined:
(1)  Ground state highly degenerate (as suggested by
the large semiclassical Hawking entropy of an extreme
Reissner-Nordstrom black hole), excited states not.
(2)  All energy levels highly degenerate,
with macroscopic energy gaps between them.
(3)  All states nondegenerate (or with low degeneracy),
separated by exponentially tiny energy gaps.
I suggest that in our world with broken supersymmetry,
this last possibility seems most plausible.
An experiment is proposed to distinguish between
these possibilities, but it would take a time
that is here calculated to be
more than about $10^{837}$ years.
}

\section{Introduction}

	What are the energy levels of a charged black hole?
For fixed electric charge $Q \gg 1$, the minimum mass of a
classical electrically charged black hole
(given by the Kerr-Newman metric, or, in the nonrotating
case, by the Reissner-Nordstrom metric) is $M=Q$.
In this paper, we shall examine the consequences
of various assumption for the levels of the excess energy
 \begin{equation}
 E \equiv M-Q.
 \label{eq:1}
 \end{equation}
(Strictly speaking, vacuum polarization is likely
to shift the minimum mass for fixed charge $Q$
slightly away from $Q$ if the theory does not
have unbroken supersymmetry, in which case
in Eq. (\ref{eq:1}), and in many equations below,
one should replace $Q$
by the minimum for the mass $M$ for the fixed charge.
However, for simplicity I shall here and henceforth
simply call this minimum mass $Q$.)

	At $E=0$, one has (classically)
an extreme-charged Reissner-Nordstrom black hole
with event horizon radius $r_+=M=Q$
and hence Bekenstein-Hawking entropy
(using Planck units throughout,
$\hbar = c = G = k = 4\pi\epsilon_0 = 1$)
 \begin{equation}
 S_0 \equiv S_H(E=0) = A/4 = \pi Q^2 \gg 1.
 \label{eq:2}
 \end{equation}
This formula for the entropy of an extreme
black hole is also supported by $D$-brane calculations
for black holes in superstring theory with unbroken
supersymmetry.
It suggests that the ground state of a
black hole with large fixed charge is highly degenerate.
On the other hand, there have been counter-arguments
\cite{GK,HHR}
suggesting that the entropy of an extreme black hole
is zero.  Which idea is correct?

	There seem to be three extreme possibilities
in our world with broken supersymmetry, though of course
intermediate possibilities also exist:
(1) Ground state highly degenerate, excited states not.
(2) All states highly degenerate, with macroscopic mass gaps.
(3) All states nondegenerate, separately microscopically.
Here I wish to examine some of the thermodynamic consequences
of these possibilities.

\section{Semiclassical Thermodynamics}

	For classical Kerr-Newman black holes with given
$Q$ and $E$ (or $M=Q+E$), the maximum Bekenstein-Hawking
entropy is obtained if the black holes have zero angular momentum
and are hence Reissner-Nordstrom black holes.
Therefore, states with zero (or at least relatively small)
angular momentum might be expected to dominate in number
for a given $Q$ and range of $E$, and so we shall focus
on such black holes.

	The classical Reissner-Nordstrom metric is
 \begin{equation}
 ds^2 = -{\Delta\over r^2}dt^2+{r^2\over\Delta}dr^2+r^2d\Omega^2,
 \label{eq:3}
 \end{equation}
where $d\Omega^2$ is the metric on the unit $S^2$ and where
 \begin{equation}
 \Delta \equiv r^2 - 2Mr + Q^2 \equiv (r-r_+)(r-r_-),
 \label{eq:4}
 \end{equation}
with
 \begin{equation}
 r_{\pm} = M \pm \sqrt{M^2-Q^2}
 \label{eq:5}
 \end{equation}
being the radii of the two horizons
($r=r_+$ at the outer or event horizon,
and $r=r_-$ at the inner or Cauchy horizon).

	In terms of $r_+$ and $r_-$, or in terms of
the event horizon radius $r_+$ and the charge $Q$,
which are two convenient parameters for describing
the Reissner-Nordstrom metrics, the black hole mass is
 \begin{equation}
 M = {1\over 2}(r_+ + r_-) = {1\over 2}r_+ + {Q^2\over 2r_+},
 \label{eq:6}
 \end{equation}
and so the energy excess over the minimum for that $Q$ is
 \begin{equation}
 E \equiv M - Q = {(r_+-Q)^2\over 2r_+}.
 \label{eq:7}
 \end{equation}
The Bekenstein-Hawking entropy is
 \begin{eqnarray}
 S &=& {1\over 4}A = \pi r_+^2 = \pi(M+\sqrt{M^2-Q^2})^2
 	\nonumber \\
   &=& \pi(Q+E+\sqrt{2QE+E^2})^2   \nonumber \\
   &\approx& \left\{ \begin{array}{ll}
   	\pi Q^2 +\pi\sqrt{8Q^3E} & \mbox{for $E \ll Q$} \\
	4\pi E^2		 & \mbox{for $E \gg Q$}
	\end{array}
	\right.
 \label{eq:8}
 \end{eqnarray}

	From these relations, one can easily derive
that the Hawking temperature is
 \begin{eqnarray}
 T_H &=& \left({\partial E\over\partial S}\right)_Q
 	= {r_+-r_-\over 4\pi r_+^2}
	= {r_+^2-Q^2\over 4\pi r_+^3}
	= {Mr_+-Q^2\over 2\pi r_+^3}
 	\nonumber \\
   &\approx& \left\{ \begin{array}{ll}
   	\sqrt{E\over 2\pi^2 Q^3} & \mbox{for $E \ll Q
		\Rightarrow T_H \sim E$ for $E \sim {1\over Q^3}$} \\
	1\over 8\pi E		 & \mbox{for $E \gg Q$}
	\end{array} \right.
 \label{eq:9}
 \end{eqnarray}
and that the heat capacity (at fixed charge) is
 \begin{equation}
 C_Q = \left({\partial E\over\partial T_H}\right)_Q
     = {2\pi r_+^2(r_+^2-Q^2)\over 3Q^2-r_+^2}.
 \label{eq:10}
 \end{equation}
The heat capacity is thus positive for $r_+ < \sqrt{3}Q$
or $M < 2Q/\sqrt{3}$ or $E < Q/(2\sqrt{3}+3) \approx 0.1547 \, Q$.

	The free energy of a Reissner-Nordstrom black hole is
 \begin{equation}
 F = E - T S = {(r_+-Q)^2\over 2r_+} - \pi T r_+^2,
 \label{eq:11}
 \end{equation}
which has no global minimum, as a function of $r_+$ (or of $E$),
for fixed $Q$ and $T$.
Therefore, even though black holes with $C_Q > 0$
are locally stable against small energy increases or decreases
from absorbing or emitting small amounts of thermal radiation,
they are globally unstable to absorbing an arbitrarily
large amount of radiation from the hypothetical infinite
heat bath.  This is a case of the general result
that in asymptotically flat spacetime
with an infinitely large heat bath at infinity,
any black hole is unstable to growing indefinitely with
negative heat capacity (once the energy gets sufficiently
larger than the charge) and entropy $S \sim 4\pi E^2$.
In other words, the canonical ensemble, weighting all
states by $e^{-E/T}$, is divergent and hence not defined
for gravitational systems in asymptotically flat spacetime,
with $T$ being the temperature at asymptotic infinity
(the only place where one can have an infinite heat bath)
\cite{H1976}.

	To define a canonical ensemble artificially,
one may introduce a negative cosmological constant,
$\Lambda \equiv -3/b^2 < 0$
\cite{HP}.
Then the metric for a charged black hole with zero
angular momentum has the same form as Eq. (\ref{eq:3}),
but now with
 \begin{equation}
 \Delta \equiv r^2 - 2Mr + Q^2 + r_+^4/b^2.
 \label{eq:11b}
 \end{equation}
The explicit formula for the event horizon radius $r_+$,
the greatest radius where $\Delta$ vanishes,
now involves the solution of a quartic polynomial
and hence is rather complicated when written in terms
of $M$, $Q$, and $b$ (or $\Lambda$),
but the explicit expressions for various quantities
remain simple when written in terms of
$r_+$ and $Q$ (and in terms of $b$, which is
assumed to be kept fixed as part of the theory).
In particular,
 \begin{equation}
 M = {1\over 2}r_+ + {Q^2\over 2r_+} + {r_+^3\over 2b^2},
 \label{eq:12}
 \end{equation}
and
 \begin{equation}
 S = \pi r_+^2 \sim \pi(2b^2M)^{2/3} \ \mbox{for $M \gg b$},
 \label{eq:13}
 \end{equation}
from which one may derive that
 \begin{equation}
 T_H = \left({\partial M\over\partial S}\right)_Q
 	= {1\over 4\pi r_+}\left({3r_+^2\over b^2}+1-{Q^2\over r_+^2}\right)
 \label{eq:14}
 \end{equation}
and
 \begin{equation}
 C_Q = \left({\partial M\over\partial T_H}\right)_Q
     = 2\pi r_+^2\left({3r_+^4+b^2r_+^2-b^2Q^2\over 3r_+^4-b^2r_+^2+3b^2Q^2}\right).
 \label{eq:15}
 \end{equation}

	The fact that the entropy does not rise
so fast as linearly with the mass for $M \gg b$ means that
with the negative cosmological constant,
the canonical ensemble is well defined
\cite{HP},
as opposed to the asymptotically flat case
(zero cosmological constant) in which
the entropy rises faster than linearly, $S \sim 4\pi M^2$.
Alternatively, the free energy (using $M$ in place of $E$,
since otherwise the expression becomes rather unwieldy) is
 \begin{equation}
 F = M - T S = {r_+^3\over 2b^2}-\pi T r_+^2+{r_+\over 2}+{Q^2\over 2r_+},
 \label{eq:16}
 \end{equation}
which is now bounded below for fixed $Q$, $T$, and $b$,
having one or two minima.

	There are two minima if $b > 6Q$,
and indeed we shall be interested in the case that $b \gg Q$,
so that the negative cosmological constant gives only
a tiny perturbation of the metric near the black hole.
Then if (but not only if) $b < 1/(\pi T)$,
the minimum at least positive $r_+$ is the global minimum.
For $Q \ll b < 1/(\pi T)$, this global minimum occurs
when $T$ is the Hawking temperature of the black hole
(with $T$ defined with respect to a timelike Killing vector
that is normalized to have unit magnitude at the
location where $g_{00} = -1$),
 \begin{equation}
 T = T_H = {1\over 4\pi r_+}
 	\left({3r_+^2\over b^2} + 1 - {Q^2\over r_+^2}\right),
 \label{eq:17}
 \end{equation}
which is very nearly that given by Eq. (\ref{eq:9})
for a Reissner-Nordstrom black hole with zero cosmological
constant, obtained by taking $b$ to infinity in Eq. (\ref{eq:17}).

	For example, the minimum that gives
$E \sim T_H \sim 1/Q^3 \ll 1$
is a global minimum of the free energy $F$ if $b \lesssim Q^3$,
which for $Q \gg 1$ (as we are assuming) allows the length
scale $b$ of the cosmological constant to be much larger
than the length scale $r_+ \approx Q$ of the nearly
extreme black hole.
To take a value of $Q$ that arises below,
$Q \sim 3 \times 10^{46} \sim 4 \times 10^8 M_{\odot}$
gives the very weak restriction
$b \lesssim Q^3 \sim 3 \times 10^{82} \, \mbox{Mpc}$
in order for this configuration to be a global
minimum of the free energy.
Therefore, for $T \ll 1/Q$, take $b \sim 1/(\pi T) \gg Q$
to stabilize the canonical ensemble but then ignore
correction terms like $r_+^2/b^2 \sim Q^2 T^2 \ll 1$
from the tiny cosmological constant.

	Some strange thermodynamic behavior occurs
when $E \lesssim T$.  Preskill {\it et al.}
\cite{PSSTW}
argued that the thermal description of a black hole
breaks down when the emission of a particle
of energy $T$ changes $T$ by a significant fraction of itself.
For a charged Reissner-Nordstrom black hole
with $E = M-Q \ll Q$, this occurs for $E \lesssim 1/Q^3$
or $T \lesssim 1/Q^3$, where $E \lesssim T$.
Such a black hole does not have sufficient energy
to emit a particle of energy $T$.

	Another argument for strange thermodynamic
behavior at $E \lesssim 1/Q^3$ came from a derivation
by Maldacena and Susskind
\cite{MSu}
of a mass gap $E_{\mbox{gap}}=1/(8Q^3)$ from string theory.
Maldacena and Strominger
\cite{MSt}
showed that it could be derived semiclassically by
implicitly assuming that the first excited state
of an extreme Reissner-Nordstrom black hole
is an extreme Kerr-Newman black hole with
angular momentum $J = 1/2$:
 \begin{equation}
 E \equiv M-Q = \sqrt{Q^2+a^2}-Q \approx {a^2\over 2Q}
 	= {J^2\over 2QM^2} \approx {J^2\over 2Q^3} = {1\over 8Q^3}.
 \label{eq:18}
 \end{equation}
Of course, this derivation fails if the lowest excited states
have $J=0$, as would occur for an ordinary chunk of solid
in the idealization of zero baryon decay.

\section{Evolving a Black Hole to Energy $E \sim 1/Q^3$}

	To get a black hole to $E = M-Q \lesssim 1/Q^3$
by emitting its excess energy, we need it to have a charge $Q$
so huge that it Hawking radiates to this point without
emitting charge.  Assume that only photons and gravitons are
massless, and that all other particles (e.g., neutrinos)
have rest masses much larger than the temperature of the hole.
Then the primary emission will be to photons with unit
angular momentum.  The three possible values of the axial
angular momentum, and the two helicities for each, then give
an energy emission rate
 \begin{equation}
 {dM\over dt} \approx
  -{6\over 2\pi}\int_0^{\infty}{P\omega d\omega\over e^{\omega/T}-1}.
 \label{eq:19}
 \end{equation}
Here $P = P(Q/M,M\omega)$ is the absorption probability
for a photon of frequency $\omega$ and unit angular momentum
by a Reissner-Nordstrom black hole of mass $M$ and charge $Q$.

	This absorption	probability can be obtained from the solution
to the radial equation
\cite{Z,Mon}
 \begin{equation}
 \left({d^2\over dr_*^2}+\omega^2-{2\Delta\over r^4}-{4Q^2\Delta\over r^6}\right)
 Z(r) = 0
 \label{eq:20}
 \end{equation}
obtained after separating variables for the photon mode, where
 \begin{equation}
 {d\over dr_*} = {\Delta\over r^2}{d\over dr} = {r^2-2Mr+Q^2\over r^2}{d\over dr}.
 \label{eq:21}
 \end{equation}
In particular, the boundary conditions for the radial mode function $Z(r)$ are
 \begin{eqnarray}
 &r_*& \rightarrow +\infty:\ \ 
 Z \sim Z_{\mbox{in}}e^{-i\omega r_*} + Z_{\mbox{out}}e^{i\omega r_*},
	\\
 &r_*& \rightarrow -\infty:\ \ 
 Z \sim Z_{\mbox{down}}e^{-i\omega r_*}.
 \label{eq:23}
 \end{eqnarray}
Then the absorption probability is
 \begin{equation}
 P = \left|{Z_{\mbox{down}}\over Z_{\mbox{in}}}\right|^2
   = 1 - \left|{Z_{\mbox{out}}\over Z_{\mbox{in}}}\right|^2.
 \label{eq:24}
 \end{equation}

	Then the goal is to evaluate $P(Q/M,M\omega)$ from
Eq. (\ref{eq:24}) and use it in Eq. (\ref{eq:19}) to find
the energy emission rate and thereby find a minimum charge $Q_*$
such that if and only if $Q > Q_*$, the time to emit energy
in uncharged particles (mainly photons) to go to $E = 1/(8Q^3) \ll 1$
is less than the time to emit a positron.
Even a single positron emission would spoil the process
of getting to $E = 1/(8Q^3) \ll 1$, since it would
reduce the black hole charge $Q$ by the positron charge
$e = \sqrt{\alpha} \approx 0.085424543$,
and reduce the black hole mass $M$
by the much smaller amount of
roughly the positron mass, $m \approx 4.18 \times 10^{-23}$
in the Planck units we are using,
so that during the positron emission, $E \equiv M - Q$
would {\it increase} by roughly $0.08542 \gg 1/(8Q^3)$.

	First, we look for an approximate solution to
the radial Eq. (\ref{eq:20}) for the frequencies that dominate
the thermal emission for a nearly extreme black hole.
These are frequencies $\omega$ that have roughly the same
order of magnitude as the Hawking temperature $T_H$.  Let
 \begin{equation}
 \epsilon \equiv {r_+-r_-\over r_+} \ll 1
 \label{eq:25}
 \end{equation}
be the tiny quantity that parametrizes how close
to extremality the black hole is.  Then
 \begin{equation}
 T_H = {\epsilon\over 4\pi r_+},
 \label{eq:26}
 \end{equation}
where, as discussed above, we are here and henceforth
neglecting the effect of the miniscule cosmological constant
needed to stabilize the canonical ensemble,
and we are also neglecting the effect of the angular momentum
of the hole, which is also expected to be relatively small
for most states of fixed $Q$ for $E = M-Q \ll Q$.
Therefore, we are considering just pure Reissner-Nordstrom
black holes that are very near their critical charge.

	Eq. (\ref{eq:26}) implies that for
frequencies $\omega$ that have roughly the same
order of magnitude as the Hawking temperature $T_H$,
the quantity $r_+\omega$ is of the order of $\epsilon$
and hence is very small.  Therefore, we shall solve
the radial Eq. (\ref{eq:20}) assuming both $\epsilon \ll 1$
and $r_+\omega \ll 1$.  This then allows
 \begin{equation}
 r_+\omega \ll \nu \equiv {r_+\omega\over\epsilon}
 = {\omega\over 4\pi T_H} \ll {1\over\epsilon}.
 \label{eq:27}
 \end{equation}
That is, the dimensionless frequency factor $\nu$
(or $4\pi\nu = \omega/T_H$) that occurs in the thermal Hawking
emission formula can range from zero (when $r_+\omega = 0$)
to a very large upper limit
(much smaller than, but within a few orders of magnitude of,
$1/\epsilon$),
thereby spanning effectively all of the thermal emission range
to give an excellent approximation to the mass emission rate
given by Eq. (\ref{eq:19}).

	To solve Eq. (\ref{eq:20}) using the approximations
of Eq. (\ref{eq:27}), split the region external to the black hole
horizon into three overlapping regions, solve Eq. (\ref{eq:20})
in each region, and match in the two overlap regions.

	In the near-horizon Region I defined by $r-r_+ \ll r_+$,
 \begin{equation}
 Z \approx Z_{\mbox{down}}{r\over r_+}\left({r-r_-\over r_+-r_-}\right)^2
   \left({r-r_+\over r-r_-}\right)^{-i\nu}
   \left[1+{4(1+i\nu)\over 1-2i\nu}\left({r-r_+\over r-r_-}\right)
     +{(1+i\nu)(1+2i\nu)\over(1-i\nu)(1-2i\nu)}\left({r-r_+\over r-r_-}\right)^2\right].
 \label{eq:28}
 \end{equation}
In the intermediate Region II defined by $r_=-r_- \ll r-r_+ \ll 1/\omega$,
 \begin{equation}
 Z \approx A{(r+2r_+)(r-r_+)^2\over r} + B{r^4(r-3r_+)\over(r-r_+)^3}.
 \label{eq:29}
 \end{equation}
And in the far-field Region III defined by $r_+ \ll r$,
 \begin{equation}
 Z \sim Z_{\mbox{in}}e^{-i\omega r_*}\left(1-{i\over \omega r_+}\right)
  + Z_{\mbox{out}}e^{i\omega r_*}\left(1+{i\over \omega r_+}\right).
 \label{eq:30}
 \end{equation}

	Then we match the solutions across Regions I and II
for $r_+-r_- \ll r-r_+ \ll r_+$, using $\epsilon \ll 1$
to guarantee the existence of this overlap region, to determine
the constants $A$ and $B$ of Eq. (\ref{eq:29}) in terms $Z_{\mbox{down}}$.
Similarly, we match the solutions across Regions II and III
for $r_+ \ll r \ll 1/\omega$, using $r_+\omega \ll 1$
to guarantee the existence of this overlap region, to determine
the constants $Z_{\mbox{in}}$ and $Z_{\mbox{out}}$ of Eq. (\ref{eq:30})
in terms of $A$ and $B$ (and hence in terms of $Z_{\mbox{down}}$
from the results of the previous overlap calculation).
The result is the absorption probability for photons of unit
angular momentum over the entire dominant part
of the Hawking emission spectrum:
 \begin{eqnarray}
 P = \left|{Z_{\mbox{down}}\over Z_{\mbox{in}}}\right|^2
   &\approx& {1\over 9}\epsilon^8\nu^4(1+\nu^2)(1+4\nu^2)
   	\nonumber \\
   &=& {1\over 9}r_+^4\omega^4(\epsilon^2+r_+^2\omega^2)(\epsilon^2+4r_+^2\omega^2).
 \label{eq:31}
 \end{eqnarray}
Again, this approximate result is valid for $\epsilon = (r_+-r_-)/r_+ \ll 1$
and for $\nu \ll 1/\epsilon$ or $\omega \ll 1/r_+ = 4\pi T_H/\epsilon$,
and within this approximation, the emission of unit angular momentum
photons overwhelmingly dominates over the emission of higher angular
momentum photons and gravitons, by several factors of the large number
$1/\epsilon$.
 
 	One can use this result to determine that the absorption cross section
of a nearly-extreme Reissner-Nordstrom black hole to photons
with $r_+\omega \ll 1$ is approximately
 \begin{equation}
 \sigma = {(2\ell+1)\pi\over\omega^2}P \approx
   {\pi\over 3}r_+^4\omega^2(\epsilon^2+r_+^2\omega^2)(\epsilon^2+4r_+^2\omega^2).
 \label{eq:32}
 \end{equation}
It is interesting that although this formula was derived only for
nearly-extreme black holes, $\epsilon \ll 1$, it gives the correct
low-frequency limit for the uncharged Schwarzschild metric,
which has $\epsilon = 1$, namely $\sigma \approx (16\pi/3)M^4\omega^2$.
On the other hand, for a precisely extreme Reissner-Nordstrom
black hole, $\epsilon = 0$ and $r_+ = Q$, we get that
the low-frequency photons ($Q\omega \ll 1$)
have $\sigma \approx (4\pi/3)Q^8\omega^6$.

	Now we can insert Eq. (\ref{eq:31}) into Eq. (\ref{eq:19})
and integrate to determine the approximate time for a black hole with
sufficiently large charge to evolve to $E\equiv M-Q = 1/(8Q^3) \ll 1$.
In particular, if the charge is sufficiently large that
$|dQ/dt| \ll |dM/dt|$ (and how large is sufficient will be discussed below),
and if $E\equiv M-Q \ll Q$, then $\epsilon \approx \sqrt{8E/Q}$
and $\omega \approx \epsilon\nu/Q \approx \nu\sqrt{8E/Q^3}$, so
 \begin{eqnarray}
 {dE\over dt} &\approx& {dM\over dt} \approx
  -{6\over 2\pi}\int_0^{\infty}{P\omega d\omega\over e^{\omega/T}-1}
  = -{3\epsilon^2\over\pi Q^2}\int_0^{\infty}{P\nu d\nu\over e^{4\pi\nu}-1}
  \nonumber \\
  &\approx& -{\epsilon^{10}\over 3\pi Q^2}
  \int_0^{\infty}{(\nu^5+5\nu^7+4\nu^9) d\nu\over e^{4\pi\nu}-1}
  = -{509\,\epsilon^{10}\over 42\,577\,920\,\pi Q^2}
  \approx -{4\,072\,E^5\over 10\,395\,\pi Q^7}.
 \label{eq:33}
 \end{eqnarray}
The time to Hawking radiate to $E=1/(8Q^3)$ is then
 \begin{equation}
 t \approx {10\,395\,\pi Q^7 \over 16\,288\,E^4}
   \approx {1\,330\,560\over 509}\pi Q^{19}
   \approx 8\,212.333\,047\,46\:Q^{19}.
 \label{eq:34}
 \end{equation}

	Next, we need to find how large $Q$ needs to be,
$Q \gtrsim Q_*$, in order that the probability be low that
the charged black hole will emit a positron during
this time and hence spoil the process of slowly evolving
to $E = 1/(8Q^3)$.  For a black hole with $1 \ll Mm \ll eQ \ll (Mm)^2$,
where $m$ and $e$ are the mass and charge of a positron,
the Hawking emission rate of positrons is well approximated
by integrating the Schwinger pair production rate
over the exterior of the black hole, where the electric
field strength is ${\mathcal{E}} = Q/r^2$:
 \begin{eqnarray}
 {dN\over dt} &\approx& \int_{r_+}^{\infty}4\pi r^2 dr
 	{e^2{\mathcal{E}}^2 \over 4\pi^3}
	\sum_{n=1}^{\infty}{1\over n^2}
		\exp{\left(-{n\pi m^2\over e{\mathcal{E}}}\right)}
  \nonumber \\
  &\approx& \int_{r_+}^{\infty}{e^2 Q^2\over \pi^2 r^2} dr
 	\exp{\left(-{\pi m^2 r^2\over eQ}\right)}
  \nonumber \\
  &\approx& {e^3 Q^3\over 2\pi^3 m^2 r_+^3}\exp{\left(-{r_+^2\over Q_0 Q}\right)}
   \approx {e^3\over 2\pi^3 m^2}\exp{\left(-{Q\over Q_0}\right)},
 \label{eq:35}
 \end{eqnarray}
where the last approximate equality assumes $\epsilon \ll 1$
so that $r_+ \approx Q$, and where
 \begin{eqnarray}
 Q_0 &\equiv& {e\over\pi m^2} = {e m_p^2\over\pi m^2}m_p^{-2}
      \approx 1.553 \times 10^{43}
     \nonumber \\
     &\approx& 1.8 \times 10^{44} e
      \approx 1.7 \times M_{\odot}
      \approx 0.84 \, s.
 \label{eq:36}
 \end{eqnarray}

	We need that this positron emission rate
be less than the reciprocal of the time to Hawking radiate
to $E=1/(8Q^3)$ that is given by Eq. (\ref{eq:34}).
This implies that we need
 \begin{equation}
 Q > Q_* = Q_0 x
 \label{eq:37}
 \end{equation}
with $x$ the solution of
 \begin{equation}
 x - 19 \ln{x} = y \equiv
 \ln{\left({665\,280 \, e^{22}\over 509\,\pi^{21}m^{40}}\right)}
 \approx 1990.
 \label{eq:38}
 \end{equation}
This implies that
 \begin{equation}
 x \approx y + 19\ln{y} + {19^2\over y}\ln{y} \approx 2136.
 \label{eq:39}
 \end{equation}
Then one gets the minimum charge $Q_*$ of a black hole
that would Hawking radiate to $E=1/(8Q^3)$
before it is likely to emit a positron
and thereby have $E \equiv M-Q$ jump upward
by much more than $1/(8Q^3)$ from the reduction
of $Q$ by the emitted positron charge $e$:
 \begin{eqnarray}
 Q_* &\approx& 3.316 \times 10^{46}
      \approx 3.88 \times 10^{47} e
      \approx 3.6 \times 10^8 M_{\odot}
      \approx 3.6 \, \mbox{AU}
     \nonumber \\
     &\approx& 5.4 \times 10^{11} \, \mbox{meters}
      \approx 1788 \, \mbox{seconds}
      \approx 0.497 \, \mbox{hour}.
 \label{eq:40}
 \end{eqnarray}
The magnitude of this last quantity, applied for the size of
a silent object that might exist in the heavens,
reminded me of a verse in the last book of the
Bible, Revelation 8:1:
``When the Lamb opened the seventh seal,
there was silence in heaven for about half an hour.''

	Now if we go back and calculate the times
given by Eq. (\ref{eq:34}) and by the reciprocal of the rate
given by Eq. (\ref{eq:35}), we find that for $Q > Q_*$
the time to Hawking radiate to $E=1/(8Q^3)$ is
 \begin{equation}
 t \approx (1.1 \times 10^{837} \mbox{yr}) \left({Q\over Q_*}\right)^{19},
 \label{eq:41}
 \end{equation}
which is less than the expected time to emit a positron,
 \begin{equation}
 t \approx (1.1 \times 10^{837} \mbox{yr})
   \exp{\left[{2136(Q-Q_*)\over Q_*}\right]}.
 \label{eq:41b}
 \end{equation}

	$Q_*$ is the charge of $6.5 \times 10^{20}$ kg of protons,
$0.88\%$ of the lunar protons, or $65\%$ of the asteroid Ceres' protons.
If superclusters of mass $M_S \sim 10^{16} M_{\odot}$ have charge
$Q_S \sim m_p M_S/e \sim 9.00 \times 10^{-19} M_S \sim 8 \times 10^{35}
\sim 2.5 \times 10^{-11} Q_*$,
we would need $\sim 4 \times 10^{10}$ such superclusters to collapse
into the black hole without the surrounding negative plasma to get
a black hole of charge $Q_*$.
If a sphere of $N_S$ such superclusters collapses,
with the $\sim N_S^{2/3}$ on the surface falling in unneutralized,
one needs $N_S \sim (4 \times 10^{10})^{3/2} = 8 \times 10^{15}$
such superclusters.  This would initially form a black hole
of mass $M \sim N_S M_S \sim 10^{32} M_{\odot} \sim 10^{70} \sim 10^{19}$ yr,
with initial charge-to-mass ratio $Q/M \sim 10^{-23}$.

	Using the results of
\cite{P1976}
for the emission of photons and gravitons
(here assumed to be the only particles of mass less than
$1/M \sim 10^{-70} \sim 10^{-42} eV$, as discussed above)
from a nearly Schwarzschild black hole,
the time to evaporate to the point where the charge
becomes important for the geometry and makes
the Hawking temperature attain its maximum value
of $1/(6\sqrt{3}\pi Q_*) \approx 9.24 \times 10^{-49}
\approx 1.31 \times 10^{-16} K \approx 1.13 \times 10^{-20} eV$
is roughly $8895\, M^3 \sim 10^{163}$ years.
Then the time to evaporate to
$E=1/(8Q^3)=1/(8Q_*^3) \approx 3.43 \times 10^{-141}
\approx 4.19 \times 10^{-113} \, eV$ (as the temperature
plummets to $1/(4\pi Q_*^3) \approx 2.18 \times 10^{-141}
\approx 3.09 \times 10^{-109} K$) would be roughly the $10^{837}$ years
discussed above.  One can see that even if the universe expands
as slowly as a matter-dominated $k=0$ Friedmann-Robertson-Walker
universe, with scale size going as the time to the 2/3 power,
the current $3 K$ microwave background would have by then cooled
below $10^{-551} K$ and so would be negligible in comparison with
the roughly $10^{-109} K$ temperature of the nearly extreme black hole.

\section{Density of States}

	Let us consider various possibilities for the density of
states (energy levels) of a nearly extreme charged black hole.
First, let us make a semiclassical estimate.  As we found above
for nearly extreme black holes, $E\equiv M-Q \ll Q$ (neglecting
the effect of the tiny cosmological constant that was introduced
to stabilize the canonical ensemble), we get that the Hawking
temperature $T_H$ (here denoted simply as $T$) is
 \begin{equation}
 T = {r_+-r_-\over 4\pi r_+^2} \approx \sqrt{E\over 2\pi^2 Q^3},
 \label{eq:42}
 \end{equation}
and the entropy is
 \begin{equation}
 S = \pi r_+^2 \approx \pi Q^2 + \sqrt{8\pi^2 Q^3 E}.
 \label{eq:43}
 \end{equation}
Expressed in terms of the temperature, this gives
 \begin{equation}
 E \approx 2 \pi^2 Q^3 T^2
 \label{eq:44}
 \end{equation}
and
 \begin{equation}
 S \approx \pi Q^2 + 4\pi^2 Q^3 T.
 \label{eq:45}
 \end{equation}

	If $E$ is reinterpreted to be
$\langle E \rangle = -d\ln{Z}/d\beta$ with $\beta = 1/T$
and partition function
 \begin{equation}
 Z(\beta) = \sum_{n=1}^{\infty}e^{-\beta E_n}
 \label{eq:46}
 \end{equation}
from energy levels $E_n$, then
 \begin{equation}
 \ln{Z} = S - \beta \langle E \rangle
 	\approx \pi Q^2 + 2\pi^2 Q^3 \beta^{-1}
 \label{eq:47}
 \end{equation}
or
 \begin{equation}
 Z \approx e^{\pi Q^2 + 2\pi^2 Q^3 T}.
 \label{eq:48}
 \end{equation}
 
	The inverse Laplace transform gives
the density of states
 \begin{equation}
 \rho(E) = {1\over 2\pi i}
 	\int_{c-i\infty}^{c+i\infty}Z(\beta)e^{\beta E} d\beta
   \approx e^{\pi Q^2}\delta(E)
     +e^{\pi Q^2}\sqrt{2\pi^2 Q^3\over E}I_1\left(\sqrt{8\pi^2 Q^3 E}\right)
     	\theta(E),
 \label{eq:49}
 \end{equation}
with $I_1$ (and $I_0$ below) being Bessel functions of imaginary argument.
The number of states $\leq E$ for $E \geq 0$ is then
 \begin{eqnarray}
 N(E) &=& \int_{-\infty}^E \rho(E')dE'
 	\approx e^{\pi Q^2}I_0\left(\sqrt{8\pi^2 Q^3 E}\right)
	\nonumber \\
      &=& e^{\pi Q^2}\left[1+{2\pi^2Q^3E\over (1!)^2}
      	+ {(2\pi^2Q^3E)^2\over (2!)^2} + \cdots\right].
 \label{eq:50}
 \end{eqnarray}

	For $1/Q^3 \ll E \ll Q$,
 \begin{equation}
 N(E) \approx {e^{\pi Q^2 + \sqrt{8\pi^2 Q^3 E}}\over 2\pi(2Q^3E)^{1/4}}
      \approx {e^{S(E)}\over 2\pi(2Q^3E)^{1/4}}.
 \label{eq:51}
 \end{equation}
If $S_0 \equiv S(E=0) = \pi Q^2$,
perhaps a good approximation for all $E \geq 0$ is that
the number of states of lower E is,
by this semiclassical estimate,
 \begin{eqnarray}
 N(E) &\approx& e^{S_0}I_0[S(E)-S_0]
 	= e^{\pi Q^2}I_0[\pi r_+^2(E) - \pi Q^2]
	\nonumber \\
      &=& e^{\pi Q^2}I_0[\pi(Q+E+\sqrt{2QE+E^2})^2 - \pi Q^2].
 \label{eq:52}
 \end{eqnarray}
This gives a huge $e^{\pi Q^2}$ degeneracy of states at $E=0$
and then a very dense quasi-continuum of states.
For $E \ll 1/(2\pi^2 Q^3)$, $N(E) \approx e^{\pi Q^2}(1+2\pi^2 Q^3 E)$,
giving a density of states
 \begin{equation}
 \rho(E) = {dN(E)\over dE} \approx e^{\pi Q^2}[\delta(E)+2\pi^2 Q^3\theta(E)].
 \label{eq:53}
 \end{equation}
The quasi-continuum of excited states has an average energy separation
 \begin{equation}
 \delta E = {1\over\rho} \approx {e^{-\pi Q^2}\over 2\pi^2 Q^3},
 \label{eq:54}
 \end{equation}
exponentially small.

	For $e^{-\pi Q^2} \ll 2\pi^2 Q^3 T \ll 1$, we get
$Z \approx e^{\pi Q^2}(1 + 2\pi^2 Q^3 T)$, which implies that
 \begin{equation}
 \langle E \rangle = -{d\ln{Z}\over d\beta} = {T^2\over Z}{dZ\over dT}
 	\approx 2\pi^2 Q^3 T^2 = (2\pi^2 Q^3 T) T \ll T.
 \label{eq:55}
 \end{equation}
Preskill {\it et al.}
\cite{PSSTW}
argue, ``The statistical treatment of the radiation
is inappropriate if the ensemble of states from which it is drawn is small.
In estimating the size of this ensemble,
we should not include the residual entropy at zero temperature,
since this is unavailable to the radiation.''

	However, contrary to this claim, there are actually
about $2\pi^2 Q^3 T e^{\pi Q^2} \gg 1$ excited states of energy $E < T$,
which is huge in absolute number.  For example, it is larger than
a googolplex ($10^{10^{100}}$) if $Q \gtrsim \sqrt{\ln{50}/\pi}\,10^{50}
\approx 8.6 \times 10^{49} \approx 2582 \, Q_*
\approx 9.37 \times 10^{11} \, M_{\odot} \approx 53$ days,
corresponding to a nearly extreme black hole with the mass of a large galaxy.
The true reason why $\langle E \rangle \ll T$ for $T \ll 1/Q^3$
is that the number of ground states is even much higher than
the enormous number of excited states with $E < T \ll 1/Q^3$,
so that the canonical ensemble is dominated by the huge
number of degenerate ground states and has only a tiny probability
for excitation.

	In particular, there are $d_0 \approx e^{\pi Q^2}$
ground states with $E=0$ for the density of states given above,
giving the total probability to be in one or another of these ground
states as $P_{\mbox{ground}} = d_0/Z \approx 1 - 2\pi^2 Q^3 T$.
Then the total probability for all excitations is
$P_{\mbox{excite}} = 1 - P_{\mbox{ground}} \approx 2\pi^2 Q^3 T \ll 1$.
Thus
 \begin{equation}
 \langle E \rangle = P_{\mbox{ground}}0
 + P_{\mbox{excite}}\langle E \ \mbox{if excited} \rangle,
 \label{eq:56}
 \end{equation}
combined with the value of $\langle E \rangle$
given by Eq. (\ref{eq:55}), implies that
 \begin{equation}
 \langle E \ \mbox{if excited} \rangle
  = {\langle E \rangle \over P_{\mbox{excite}}}
  \approx T,
 \label{eq:57}
 \end{equation}
which is precisely what one would expect
if there were a very dense uniform distribution of states,
when $T$ is greater than the average separation between
the levels.

	If the lowest excited state is nondegenerate
and has $E = \delta E \sim e^{-\pi Q^2}/(2\pi^2 Q^3)$,
which is the average separation between energy levels
for the density of states given by Eq. (\ref{eq:53}),
and if the next excited state has $E = (1+f)\delta E$,
etc., then for $T \ll f\delta E$,
$P_{\mbox{excite}} \approx e^{-\pi Q^2 - \delta E/T}$ and
$\langle E \rangle \approx (\delta E)e^{-\pi Q^2 - \delta E/T}$.
Thus it is only for $T \lesssim 1/(\delta E)$
that the absolute scarcity of excited states
with $E < T$ has a significant effect in changing the form of
$\langle E \rangle$;
for $1/(\delta E) \ll T \lesssim 1/Q^3$,
the number of excited states with $E < T$ is huge in absolute number,
and it is only the fact that it is less than the even larger
number of degenerate ground states that makes $\langle E \rangle \ll T$.

	Thus the semiclassical estimate gives
a ground state degeneracy of $d_0 \approx e^{\pi Q^2}$
and then a quasi-continuum with an exponentially large density of states
$\rho(E) \approx 2\pi^2 Q^3 e^{\pi Q^2}$ for $E \lesssim 1/Q^3$.

	One alternative for the energy states
of a nearly extreme charged black hole is that there
is a whole series of highly degenerate levels,
separated by macroscopic mass gaps.
For example, the classical formula for extreme
black holes with angular momentum $J = n/2$
for nonnegative integers $n$ is
 \begin{equation}
 E_n = \sqrt{{1\over 2}[Q^2+\sqrt{Q^4+4J^2}]}-Q
 = Q \left\{\sqrt{{1\over 2} + {1\over 2}\sqrt{1+{n^2\over Q^4}}}-1\right\}
 = {n^2\over 8Q^3} - {5n^4\over 128 Q^7} + \cdots .
 \label{eq:58}
 \end{equation}
One might postulate that each of these levels
has degeneracy
 \begin{equation}
 d_n \approx e^{S(E,J)} = e^{\pi r_+^2}
 = e^{\pi(Q+E_n)^2} \approx e^{\pi Q^2}\left(1+{\pi n^2\over 4Q^2}\right),
 \label{eq:59}
 \end{equation}
possibly multiplied by some function of $Q$ and $n$
that is not exponentially large in $Q^2$.

	If $E_n \approx E_1 n^a$ and $d_n \propto n^b$
for some constants $a$ and $b$, then for
$T \equiv 1/\beta \gg E_1$,
 \begin{equation}
 Z = \sum_{n=0}^{\infty} d_n e^{-\beta E_n}
 \propto \int_0^{\infty} dn \, n^b \, e^{-\beta E_1 n^a}
 \propto \beta^{-(1+b)/a},
 \label{eq:60}
 \end{equation}
which then implies that
 \begin{equation}
 \langle E \rangle \approx {1+b\over a} T.
 \label{eq:61}
 \end{equation}
For example, the alternative above, in which
$E_n \approx n^2/(8Q^3)$
and $d_n$ is approximately independent of $n$,
corresponds to $a=2$ and $b=0$ and hence gives
$\langle E \rangle \approx T/2$
for $E_1 \approx 1/(8Q^3) \ll T \ll 1/Q$.
On the other hand, for $T \ll E_1$,
$Z \approx d_0 + d_1 e^{-\beta E_1}$,
which implies that
$\langle E \rangle \approx (d_1/d_0)e^{-E_1/T}$,
which is exponentially suppressed when there
are no states in an initial mass gap.

	There is an opposite alternative for the energy states
of a nearly extreme charged black hole that to me seems
more plausible in our real world with broken supersymmetry.
This alternative is that there is no huge degeneracy anywhere,
even for the ground state, but instead only an exponentially
dense quasi-continuum of states.

	For example, if one takes the semiclassical estimate
above and then changes the number of ground states from
the value above of $d_0 \approx e^{\pi Q^2}$ to $d_0 = 1$,
then one just subtracts roughly $e^{\pi Q^2}-1$ from the
semiclassical estimate for the partition function.
This also reduces by roughly $e^{\pi Q^2}-1$
the number of states of energy $\leq E$
(for $E \geq 0$; for $E < 0$ there are no states)
from the approximate value given by Eq. (\ref{eq:52})
to what now becomes, after inserting the operator
Int to take the integer part of what follows,
 \begin{eqnarray}
 N(E) &\approx& \mbox{Int} \left\{e^{S_0}[I_0(S(E)-S_0)-1]+1\right\}
 	\nonumber \\
      &=& \mbox{Int} \left\{e^{\pi Q^2}[I_0(\pi r_+^2(E)-\pi Q^2)-1]+1\right\}
	\nonumber \\
      &=& \mbox{Int} \left\{e^{\pi Q^2}[I_0(\pi(Q+E+\sqrt{2QE+E^2})^2-\pi Q^2)-1]+1\right\}
        \nonumber \\
      &\approx& \mbox{Int} \left\{e^{\pi Q^2}[I_0(\sqrt{8\pi^2Q^3E})-1]+1\right\}
        \nonumber \\
      &\approx& \mbox{Int} \left\{2\pi^2 Q^3 e^{\pi Q^2} E + 1\right\}.
 \label{eq:62}
 \end{eqnarray}
Here the penultimate expression on the right hand side (after the second
$\approx$ sign) applies for $E \ll Q$, any nearly extreme charged black hole,
and the final expression on the right hand side applies for
$E \ll 1/Q^3 \ll 1$, where the density of states is nearly uniform,
with average spacing given by Eq. (\ref{eq:52}).

	If the temperature is much greater than this average spacing,
$e^{-\pi Q^2}/(2\pi^2 Q^3)$, but much less than the maximum temperature
of a Reissner-Nordstrom black hole of fixed charge $Q$,
$1/(6\sqrt{3}\pi Q)$, so that then the canonical ensemble
is dominated by a large number of excited states that
are all nearly extreme black holes
(using a tiny cosmological constant to suppress the contribution
of the much larger nearly Schwarzschild-anti-deSitter black holes
that would have similar or lower temperatures if the cosmological
constant were zero), then the partition function has the form
given by subtracting roughly $e^{\pi Q^2}$ from that given by Eq. (\ref{eq:48}):
 \begin{equation}
 Z \approx e^{\pi Q^2 + 2\pi^2 Q^3 T} - e^{\pi Q^2}.
 \label{eq:63}
 \end{equation}
In this temperature range, one thus gets
 \begin{eqnarray}
 \langle E \rangle &=& {T^2\over Z}{dZ\over dT}
	\approx {2\pi^2 Q^3 T^2 \over 1 - e^{-2\pi^2 Q^3 T}}
   \nonumber \\
        &\approx& \left\{ \begin{array}{ll}
   	T & \mbox{for $e^{-\pi Q^2} \ll 2\pi^2 Q^3 T \ll 1$} \\
	2\pi^2 Q^3 T^2  & \mbox{for $1 \ll 2\pi^2 Q^3 T$}
	\end{array} \right.
 \label{eq:64}
 \end{eqnarray}
Furthermore, for a fixed temperature in this range
(and with a tiny cosmological constant to stabilize
the canonical ensemble, as discussed above), the von Neumann entropy
of the thermodynamic canonical ensemble with
the presently-assumed density of states is
 \begin{eqnarray}
 S &=& \ln{Z} + {\langle E \rangle \over T}
	\approx \pi Q^2 + 2\pi^2 Q^3 T
	\left({2 - e^{-2\pi^2 Q^3 T} \over 1 - e^{-2\pi^2 Q^3 T}}\right)
	+ \ln{(1 - e^{-2\pi^2 Q^3 T})}
   \nonumber \\
        &\approx& \left\{ \begin{array}{ll}
   	\pi Q^2 + 1 + \ln{(2\pi^2 Q^3 T)}
		& \mbox{for $e^{-\pi Q^2} \ll 2\pi^2 Q^3 T \ll 1$} \\
	\pi Q^2 + 4\pi^2 Q^3 T & \mbox{for $1 \ll 2\pi^2 Q^3 T$}
	\end{array} \right.
 \label{eq:65}
 \end{eqnarray}

	On the other hand, for a temperature comparable to or smaller than
the spacing between energy levels, $T \lesssim e^{-\pi Q^2}/(2\pi^2 Q^3)$,
the discreteness of the energy levels is important.
If the levels are indeed evenly spaced with a separation of
$e^{-\pi Q^2}/(2\pi^2 Q^3)$ for $E \lesssim 1/Q^3$,
then for all $T \lesssim 1/Q^3$ one gets the thermodynamics
of a single harmonic oscillator of frequency given by the energy spacing,
 \begin{equation}
 Z \approx \left[1-\exp{\left(-{1\over 2\pi^2 Q^3 e^{\pi Q^2}T}\right)}\right]^{-1},
 \label{eq:66}
 \end{equation}
 \begin{equation}
 \langle E \rangle \approx \left\{2\pi^2 Q^3 e^{\pi Q^2}
 	\left[\exp{\left({1\over 2\pi^2 Q^3 e^{\pi Q^2}T}\right)}-1\right]\right\}^{-1},
 \label{eq:67}
 \end{equation}
 \begin{equation}
 S \approx \left\{2\pi^2 Q^3 e^{\pi Q^2} T
 	\left[\exp{\left({1\over 2\pi^2 Q^3 e^{\pi Q^2}T}\right)}-1\right]\right\}^{-1}
     -\ln{\left[1-\exp{\left(-{1\over 2\pi^2 Q^3 e^{\pi Q^2}T}\right)}\right]}.
 \label{eq:68}
 \end{equation}
The second and third of these reduce to the expressions
given in Eqs. (\ref{eq:64}) and (\ref{eq:65}) for
$e^{-\pi Q^2} \ll 2\pi^2 Q^3 T \ll 1$,
but for $T \ll e^{-\pi Q^2}/(2\pi^2 Q^3)$,
$\langle E \rangle$ and $S$ (as well as $Z-1$)
are exponentially suppressed.

	These expressions show that if the ground state
of a black hole of fixed charge $Q$ is not degenerate
(or has a degeneracy much less than the exponential
of one-quarter its area of $4\pi Q^2$),
but the density of excited states is roughly that given
by the semiclassical estimate,
then for temperatures $T$ and excess energies $E$
(i.e., energies above that of the ground state
energy, which is roughly $Q$ for the fixed charge $Q$)
greater than about $1/Q^3$,
the entropy and other thermodynamic properties
are very near what one would calculate for a classical
black hole with mass given by the charge plus the expectation
value of the excitation energy.
However, for temperatures and excess energies less than about $1/Q^3$,
then the absence of a huge ground-state degeneracy changes
the thermodynamic values from what a classical calculation
would indicate. In particular, as one lowers the temperature,
the thermodynamic entropy is reduced below one-quarter the area
of the corresponding classical Reissner-Nordstrom black hole
with the same classical temperature or same energy
as that of the thermal expectation value.
Indeed, when the temperature and excess energy expectation value
are reduced to zero, the entropy reduces to the logarithm
of the number of degenerate ground states, which is zero
if the ground state is not degenerate.

	These deduced thermodynamic properties
at very low temperatures and energies are different from
what a na\"{\i}ve classical calculation would give,
but they seem to be a plausible possibility of what
happens for charged gravitational systems.
Of course, the simple modification of the semiclassical
calculation by dropping the degeneracy of the ground state
is also na\"{\i}ve, so its detailed predictions are
not likely to be correct (e.g., a fairly uniform
density of states all the way down to a nondegenerate
ground state), but they seem to be a reasonable
zeroth-order guess and plausibly are qualitatively correct.

	How might one be able to tell in principle
whether the ground state of a charged black hole
has a huge degeneracy or not?  One way is to examine
the time for the hole to become extreme.

	If the ground state has a huge degeneracy $\sim e^{\pi Q^2}$,
comparable to the number of excited states with excitation energies
up to about $1/Q^3$, then after the black hole radiates down to
$E \sim 1/Q^3$, a reasonable fraction of the lower-energy states
are ground states, so it seems plausible that the hole may go to
$E=0$ in one more transition with reasonable probability.
Then the expected time to go to one of the ground states would be comparable
to (e.g., say a factor of two greater than) the expected time
to go to some energy of the order of $1/Q^3$, which we estimated above
to be about $10^4 \, Q^{19}$ and greater than
the expected time to emit a positron for
$Q \gtrsim Q_* \sim 4 \times 10^8 M_{\odot}$,
which implies a time equal to or greater than about $10^{837}$ years
for a black hole of charge $Q \gtrsim Q_*$ to become extreme
and stop emitting (except for emitting positrons, which we found occurs
at a slower rate that is suppressed by an exponential of $Q/Q_*$).

	However, if the ground state of a charged black hole is
has only a small degeneracy, say smaller by a factor of roughly
$e^{\pi Q^2}$, then one might expect that the time to decay through
all the excited states and reach one of the small number (e.g., one) of
ground states might be larger by this same factor (assuming that $Q$
stayed constant).  But a time of order $e^{\pi Q^2}$,
whose logarithm grows quadratically with $Q$,
is always larger than the time to emit a positron,
which goes as $e^{Q/Q_*}$, whose logarithm rises only linearly
with $Q$.  So if this is indeed the case,
one would not expect ever to reach a ground state for nonzero charge;
the charge would almost always decrease from any fixed value
before one reached the ground state for that charge.

	Therefore, to tell whether a charged black hole
has a ground state degeneracy $\sim e^{\pi Q^2}$
for $Q \gg Q_* \sim 4 \times 10^8 M_{\odot}$, just wait several times
$10^4 \, Q^{19} \sim 10^{837} \mbox{yr} (Q/Q_*)^{19}$
and see whether the black hole has stopped radiating.
Unfortunately, this is slightly longer than
the time over which a Ph.D. student generally hopes
to complete his or her thesis research and graduate.

	This work was supported in part by the
Natural Sciences and Engineering Council of Canada.




\begin{thebibliography}{99}

\bibitem{GK}  G.W.\ Gibbons and R.E.\ Kallosh,
Physical Review {\bf D51} (1995) 2839.

\bibitem{HHR}  S.W.\ Hawking, G.T.\ Horowitz, and S.F.\ Ross,
Physical Review {\bf D51} (1995) 4302.

\bibitem{H1976}  S.W.\ Hawking, Physical Review {\bf D13} (1976) 191.

\bibitem{HP}  S.W.\ Hawking and D.N.\ Page,
Communications in Mathematical Physics {\bf 87} (1983) 577.

\bibitem{PSSTW}  J.\ Preskill, P.\ Schwarz, A.\ Shapere, S.\ Trivedi, and F.\ Wilczek,
Modern Physics Letters {\bf A6} (1991) 2353.

\bibitem{MSu}  J.M.\ Maldacena and L.\ Susskind,
Nuclear Physics {\bf B475} (1996) 679. 

\bibitem{MSt}  J.M.\ Maldacena and A.\ Strominger,
Physical Review {\bf D56} (1997) 4975.

\bibitem{Z}  F.J.\ Zerilli, Physical Review {\bf 9} (1974) 860.

\bibitem{Mon}  V.\ Moncrief, Physical Review {\bf 9} (1974) 2707,
{\bf 10} (1974) 1057, {\bf 12} (1975) 1526.

\bibitem{P1976}  D.N.\ Page, Physical Review {\bf D13} (1976) 198.

\end{thebibliography}
\end{document}